# X-ray fluorescence induced by standing waves in the grazing incidence and grazing exit modes: study of the Mg-Co-Zr system

*Yuchun Tu[1,2,3,+], Yanyan Yuan[1,2,++], Karine Le Guen[1,2], Jean-Michel André[1,2], Jingtao Zhu[3], Zhanshan Wang[3], Françoise Bridou[4], Angelo Giglia[5], Philippe Jonnard[1,2,*]*

[1]Sorbonne Universités, UPMC Univ Paris 06, Laboratoire de Chimie Physique - MatièreetRayonnement,11 rue Pierre et Marie Curie, F-75231 Paris cedex 05, France

[2]CNRS UMR 7614, Laboratoire de Chimie Physique - MatièreetRayonnement, 11 rue Pierre et Marie Curie, F-75231 Paris cedex 05, France

[3] MOE Key Laboratory of Advanced Micro-structured Materials,School of Physics Science and Engineering,Tongji University, Shanghai 200092, China

[4]Laboratoire Charles Fabry, Institutd'Optique, CNRS UMR 8501, Université Paris Sud 11, 2 avenue Augustin Fresnel, F-91127 PalaiseauCedex, France

[5] CNR-IstitutoOfficinaMateriali, I-34149 Trieste, Italy

**\* Corresponding Author:** philippe.jonnard@upmc.fr
+ Dr.YuchnTu: on leave to the Shanghai Institute of Laser Plasma, China.
++Dr.YanyanYuan: on leave to the Jiangsu University of Science and Technology, China.

ABSTRACT. We present the characterization of Mg-Co-Zr tri-layer stacks by using x-ray fluorescence induced by x-ray standing waves, both in the grazing incidence (GI) and grazing exit (GE) modes. The introduction of a slit in the direction of the detector improves the angular resolution by a factor 2 and significantly the sensitivity of the technique for the chemical characterization of the buried interfaces. By observing the intensity variations of the Mg K$\alpha$ and Co L$\alpha$ characteristic emissions as a function of the incident (GI mode) or detection (GE mode) angle, we show that the interfaces of the Si/[Mg/Co/Zr]$_{x30}$ multilayer are abrupt, whereas in the Si/[Mg/Zr/Co]$_{x30}$ multilayer a strong intermixing occurs at the Co-on-Zr interfaces. The explanation of this opposite behaviour of the Co-on-Zr and Zr-on-Co interfacesis given by the calculation of the mixing enthalpies of the Co-Mg, Co-Zr and Mg-Zr systems, which shows that the Co-Zr system presents anegative value and the two others positive values. Together with the difference of the surface free energies of Zr and Co, this leads us to consider the Mg/Zr/Co system as aMg/Co$_x$Zr$_y$bi-layer stack, with x/y estimated around 3.5.

1. INTRODUCTION

Periodic multilayers alternating two or more layers of nanometer thickness can be used to diffract radiation in the x-ray and extreme ultra-violet ranges(Attwood, 2000). Thus, they are used as optical components for numerous applications in photolithography, x-ray microscopy, x-ray spectroscopy, in space telescopes or on synchrotron beamlines. However, the optical performance of such multilayer stacks greatly depends on the quality of their interfaces. So it is important to characterize them, which is often done by x-ray reflectivity in the hard x-ray range or at the application wavelength or by transmission electron microscopy.

Recently we have demonstrated(Jonnard et al., 2014)that x-ray fluorescence (XRF) generated by x-ray standing waves (XSW), combining both grazing incidence (GI) and grazing exit (GE) modes, is an efficient meansfor the characterization of such stacks. Indeed, the standing wave generated by the incident radiation (GI mode) or characteristic emission (GE mode) has the same period as the multilayer. Then, by rotating the sample in an angular range centered around the Bragg angle of the incident or emitted radiation, it is possible to move the nodes and anti-nodes of the electric field at specific positions within the stack, an interface or the center of a layer, and thusto locate the origin of the generated x-ray signal with great depth sensitivity(Bedzyk & Libera, 2013). This technique is related to the technique of x-ray standing wave at grazing incidence and exit(Sakata & Jach, 2013).

We apply GI- and GE-XRF to the Mg-Co-Zrtri-layer system whose period thickness, around 9.5 nm, was designed to have constructive interferences and Bragg peak at not extreme grazing incidence. Two tri-layers, Si/[Mg/Co/Zr]$_{x30}$and Si/[Mg/Zr/Co]$_{x30}$, weredeposited by magnetron sputtering, where the order of the layers in the stack is different. This kind of multilayers had already been studied for optical applications around the 25 nm spectral range(Le Guen et al., 2011a; b). With respect to the present multilayers, the Co and Zr thicknesses were the same whereas Mg layers were much thicker. They have been thoroughly characterized by x-ray reflectivity in the hard and soft x-ray ranges, x-ray emission spectroscopy, nuclear magnetic resonance spectroscopy, secondary ion mass spectrometry and also transmission electron microscopy(Le Guen et al., 2011a; b; Zhu et al., 2011; Jonnard et al., 2013).

This paper is organized as follows. Firstly, in the experimental section we indicate how the samples are prepared and recall some details of the GI- and GE-XRF experiments. We present the improvement of the angular resolution in the GE mode with respect to our previous study of the Co/Mg bi-layer system, The way simulations are obtained and handledfor comparison with the experimental curves is then described. Finally, we present and discuss comparatively the results for both tri-layer systems, obtained in the GI and GE modes giving the angular variations of the intensity of the Mg K$\alpha$(2p-1s transition) and Co L$\alpha$(3d-2p$_{3/2}$ transition) characteristic emissions.

## 2. EXPERIMENTAL METHODS

### 2.1 Samples

The deposition of the samples was done by magnetron sputtering ontosilicon wafers used as substrates. After deposition, a 3.5 nm-thick boron carbide (B$_4$C) thin layer was added as a capping layer to prevent samples from oxidation. Different multilayers were deposited on the basis of the Si/[Mg(5.45 nm)/Co(2.45 nm)]$_{30}$/B$_4$C (3.5 nm) bi-layer system.The Mg/Co bi-layer system had already been studied by GI- and GE-XRF previously(Jonnard et al., 2014) and is used here with the purpose to demonstrate the improvement of the angular resolution and to reveal that the combination of both GI and GE modes is powerful for the characterization of buried interfaces of already very well studied systems. The two tri-layer samples are :

- Si / [Mg(5.45 nm) / Co(2.45 nm) / Zr(1.50 nm)]$_{30}$ / B$_4$C(3.5 nm), noted Mg/Co/Zr;
- Si / [Mg(5.45 nm)/ Zr(1.50 nm) / Co(2.45 nm)]$_{30}$ / B$_4$C(3.5 nm), noted Mg/Zr/Co.

With this notation the layers are written in the order of their deposition.

Following deposition, all the samples were characterized by grazing incident x-ray reflectivity at the Cu K$\alpha$ wavelength(0.154 nm). The thickness, roughness and density of the each layer were determined by fitting the reflectivity curves using the designed one as a model of the stack. The results are givenin Table 1 for the tri-layer systems.The stack parameters are close to the designed ones. The roughness, or interface width is limited to around 0.5 nm.

**Table 1.** Parameters of the tri-layersystems as deduced from the fits of their reflectivity curves. For each layer, are indicated its thickness (nm), density(g.cm$^{-3}$) and interface width (nm).

| Sample    | Period(nm) | Mg             | Co             | Zr             |
|-----------|------------|----------------|----------------|----------------|
| Mg/Co/Zr  | 9.45       | 5.22 / 1.6 / 0.5 | 2.61 / 8.8 / 0.5 | 1.62 / 6.5 / 0.6 |
| Mg/Zr/Co  | 9.57       | 5.49 / 1.6 / 0.5 | 2.35 /8.8 / 0.5  | 1.73 / 6.5 / 0.6 |

### 2.2 Schemes of the experiments

The experimental details had already been given(Jonnard et al., 2014) andhere we only recall the main characteristics of our two experimental procedures, both performed on the same samples. Experiments were performed at the BEAR beamline of the ELETTRA synchrotron radiation facility(Nannarone et al., 2004). Prior to the XSW experiments, we obtain an XRF spectrum of one sample with a silicon drift detector (SDD) to determine in which spectral region the fluorescence emission of an element, Co L$\alpha$or Mg K$\alpha$ in our case, should be integrated. The incident photon energy to excite the Co Lα emission was 807.6 eV; it was 1332 eV to excite the Mg Kα emission.

In the GI-XRF mode, the intensity of an emission is measured as a function of the glancing angle *i*, *i.e.* the angle between the synchrotron beam and the sample surface, for angles close to the Bragg angle calculated from the period of the sample and the wavelength of the incident radiation. In the GE-XRF mode, the intensity of an emission is measured as a function of the take-off angle of emission *d*, *i.e.* the angle between the detector and the sample surface, for angles close to the Bragg angle calculated from the period of the sample and the wavelength of the characteristic emitted radiation. This is illustrated in Figure 1. In our case there is a fixed angle of 60° between the directions of the incident and the detected radiations. It results from this mechanical constraint that $d = 120 - i$ (angles en degrees). In both modes, the SDD is located in the incidence plane, *i.e.* the plane defined by the incident beam and the normal to the sample surface.

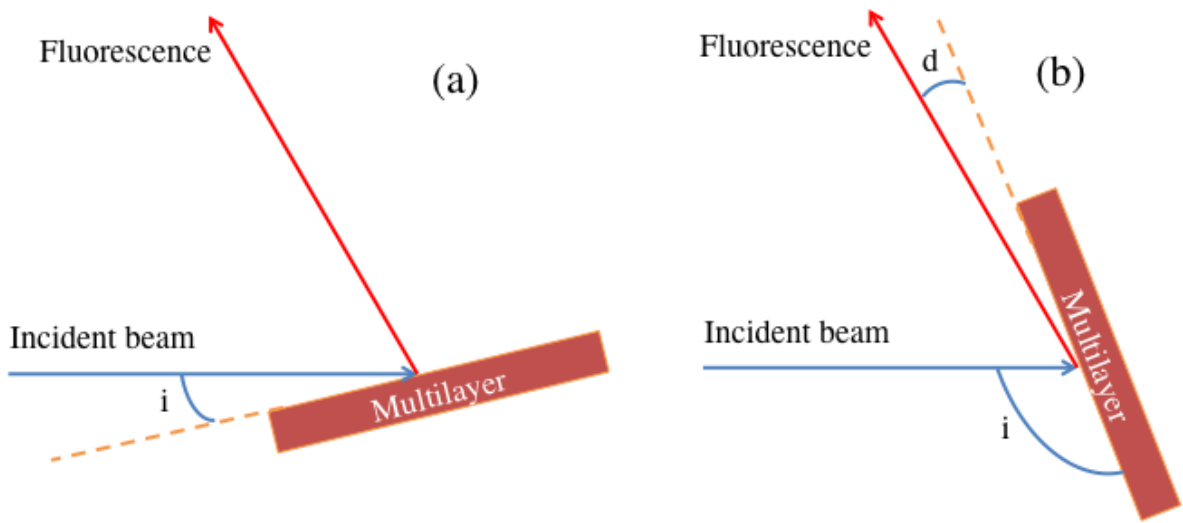

**Figure 1.** Experimental configurations for the GI-XRF (a) and GE-XRF (b) experiments. *i* and *d* are the glancing angle and glancing take-off angle respectively.

## 2.3 Angular resolution in GE-XRF mode

In the GI-XRF the angular resolution is governed by the divergence of the incident beam, that is to say of the synchrotron radiation beam. It is quite small, 0.2° in our experimental conditions. In the GE-XRF the angular resolution is governed by the aperture of the detector. Taking into account the aperture of the SDD, 5 mm, and its distance from the sample, 300 mm, leads to a 0.9° angular aperture. As shown in Figure 2, to improve the angular resolution, we can insert two slits with width respectively of 1.0 or 0.5 mm, at 140 mm from the sample and 160 mm from the SDD. These configurations correspond to angular acceptances of 0.4 and 0.2°, respectively. The rotation of the slit-holder around the azimuthal axis allows putting the desired slit in the incidence plane.

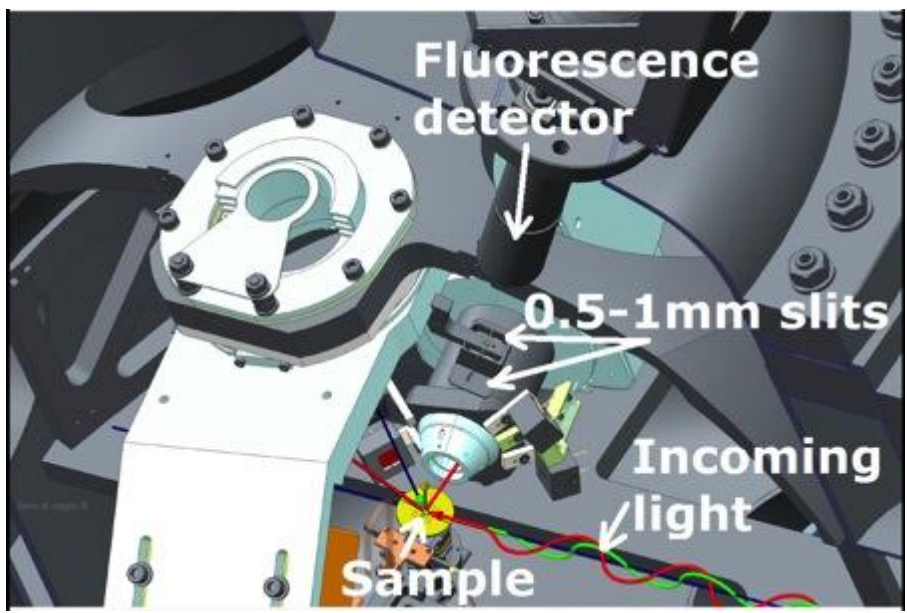

**Figure 2.** Experimental setup: the slit holder can be rotated to put one of the two narrow slits in the axis going from the sample to the detector.

We present in Figure 3, the GE-XRF curves showing the variations of the Mg Kα intensity of the Co/Mg sample with the three possible configurations to check the angular resolution: without slit, and slit of 1.0 mm or 0.5 mm. In all three cases, the main feature around 3°, relative to the first order diffraction of the emitted radiation, is clearly observed. However, the feature regarding the second order diffraction of the emitted radiation, located around 6°, is hardly seen without a slit and is better distinguished when the slit width decreases. This is due to the improvement of the angular resolution, as can be seen on the normalized curves (Fig. 3b). We estimate to have gained a factor 2 on the resolution when going from the configuration with no slit to the one with the 0.5 mm slit. This was estimated from the angular distance between the maximum and the dip of the first order feature. Let us note that the improvement of the resolution is detrimental the collected intensity, which decreases by a factor 3, as can be seen in the raw data (Fig. 3a). However, owing to the large incident flux delivered by the synchrotron, we choose in the following to work in the configuration with the narrowest slit.

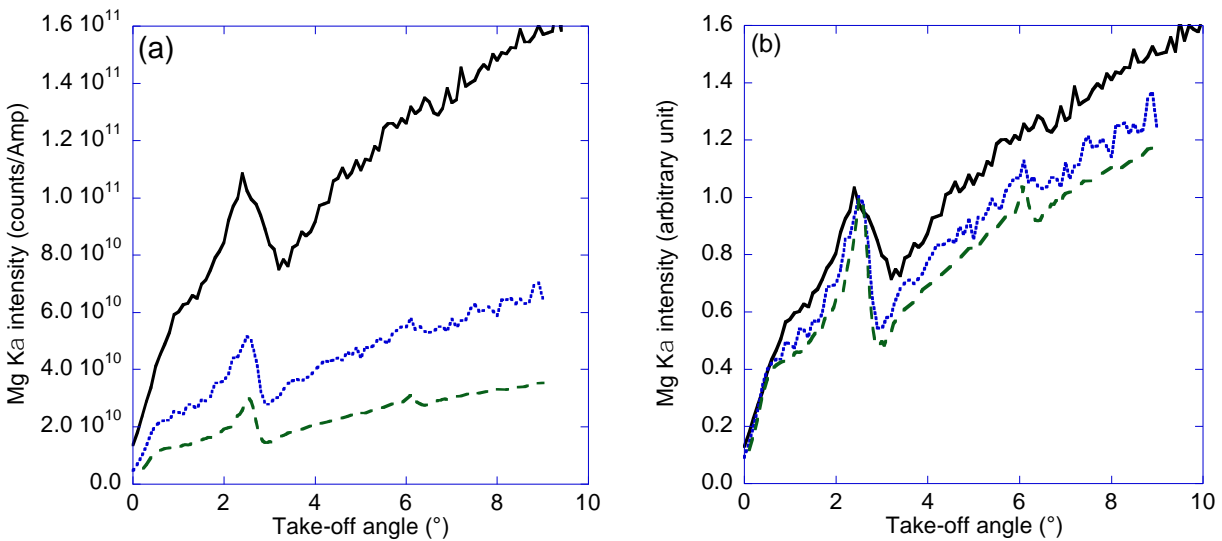

**Figure 3.** GE-XRF of the Mg Kα emission of the Co/Mg sample: (a) raw data; (b) data normalized to the maximum of the first diffraction order feature. Measurement with no detection slit (solid line), with a 1.0 mm slit (dotted line) and with a 0.5 mm slit (dashed line).

### 2.4 Fit of the experimental curves

The fit of the experiment relies on simulations of the intensity of the characteristic fluorescence radiation. The model having been already presented(Chauvineau & Bridou, 1996; Jonnard et al., 2014), we recall only its main points. First, the intensity of the exciting electromagnetic field at a given depth in the stack and below a given glancing angle $i$, is calculated. This is done from the recurrent formalism used to calculate the optical properties of stacks from the optical constants and thicknesses of its various layers. The intensity generated at the same depth from an element of a given concentration is proportional to the square of the electric field and to the concentration. Then, the fluorescence intensity arriving below thetake-off angle $d$ from the source to be located is calculated by applying the reciprocity theorem(von Laue, 1935; James, 1962). Let us note that the chosen formalism can be applied to both GI and GE modes. Thus, the fits are performed so that a best agreement is obtained for both GI and GE curves at the same time. This imposes a strict constraint on the fitted parameters. They are the same as the ones determined from the reflectivity measurements (thickness, roughness and density of the layers), which are generally used as input values. The roughness of the various layers is introduced and considered fixed at the value given by the reflectivity measurement. The boron carbide capping layer is also considered with its nominal thickness.

Comparing properly experimental and simulated curves requiresto take into account the effect of the beam imprint on the sample. Indeed, contrary to the case of reflectivity measurements where the specular geometry ensures that the imprint of the incident and reflected beams is the same, this is not the case for GE and GI measurements owing to the asymmetric geometry. Thus in the GI case, a *1/sin(i)* factor(Li et al., 2012) is introduced. For the GE mode, no geometrical factor was taken into account since this correction is significant only for the small glancing angles. The simulated curves are then normalized on one point of the experimental curve.

## 3. RESULTS AND DISCUSSION

For both Mg K$\alpha$ and Co L$\alpha$ emissions, we show in Figures 4 and 5the GI- and GE-XRF curves of the Mg/Co/Zr and Mg/Zr/Co multilayers, respectively. The modulationof the intensity is clearly observed on each curve at an angleclose to the one calculated from the Bragg law at the first and seconddiffraction orders.It can be seen that the amplitude of the observed features are quite similar in the case of the Mg K$\alpha$ emission of both samples, whereas for the Co L$\alpha$ emission, the featuresexhibit a higher contrast for Mg/Co/Zr(Fig. 4c and 4d) with respect to the ones of Mg/Zr/Co (Fig. 5c and 5d). This is valid for both GI and GEmodes and can be seen in Figures 6(a) and 6(b) where the experimental XSW curves of the Mg/Co/Zr and Mg/Zr/Co multilayers are compared.

The simulation of the Mg K$\alpha$ emission of Mg/Co/Zr and Mg/Zr/Co multilayers is done with the parameters deduced from the reflectivity measurements, thus considering a tri-layer stack. We can deduce that in these multilayers, the interfaces involving the Mg layers, Mg-Co and Mg-Zr, are abrupt. This is in agreement with previous results(Le Guen et al., 2011a; b; Zhu et al., 2011) of nuclear magnetic resonance and x-ray emission spectroscopies giving the chemical state of the Co and Mg atoms in the stack, respectively.

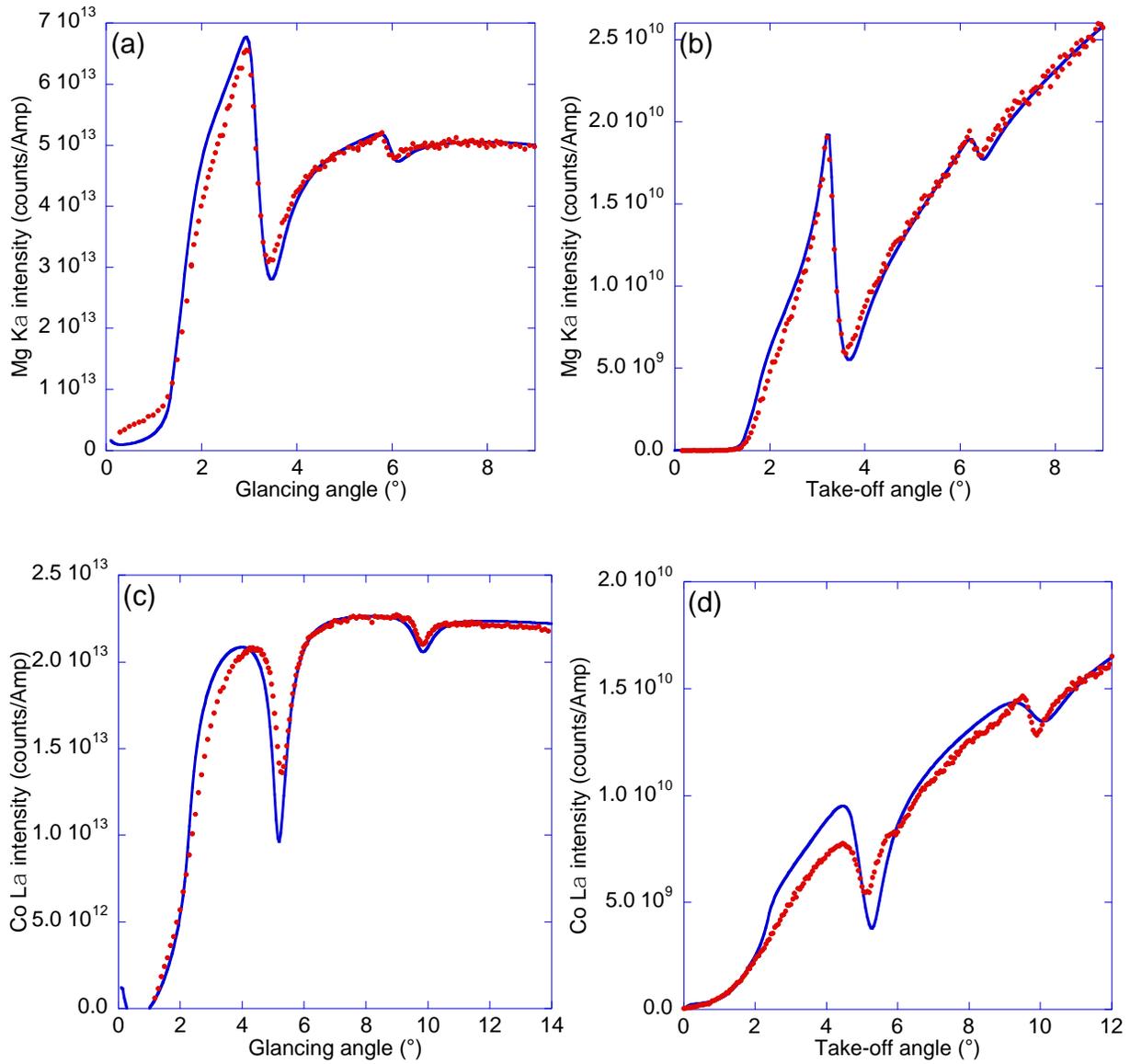

**Figure 4.** GI-XRF (a) and (c) and GE-XRF (b) and (d) curves for the Mg Kα (a) and (b) and Co Lα (c) and (d) emissions of the Mg/Co/Zr multilayer: dots, experiment; solid line: fit.

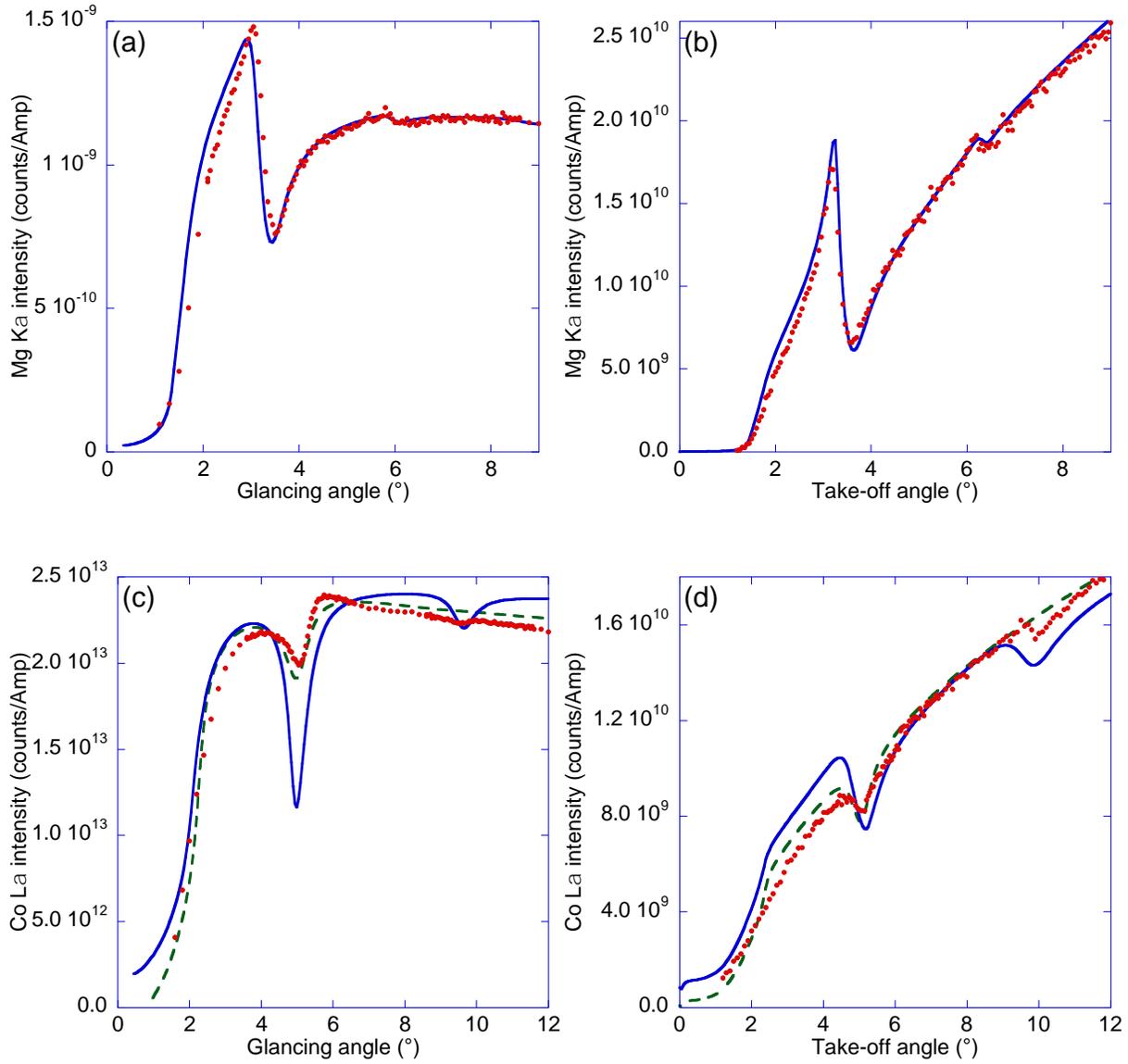

**Figure 5.** GI-XRF (a) and (c) and GE-XRF (b) and (d) curves for the Mg Kα (a) and (b) and Co Lα (c) and (d) emissions of the Mg/Zr/Co multilayer: dots, experiment; solid line: fit with a tri-layer stack; dashed line: fit with a bi-layer stack.

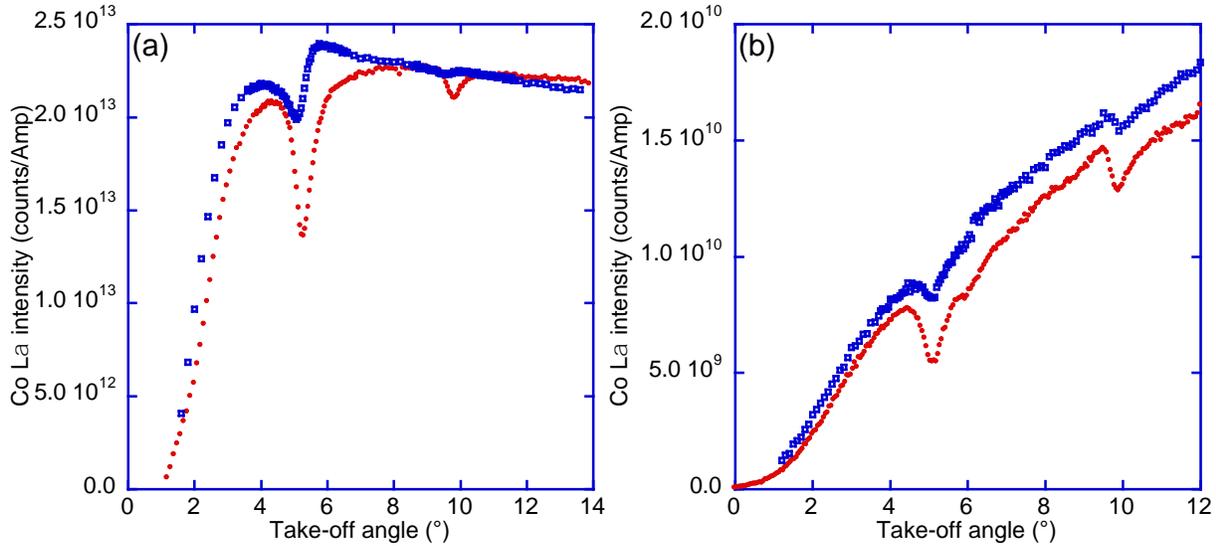

**Figure 6.** GI-XRF (a) and GE-XRF (b) curves for the Co Lα emission of the Mg/Co/Zr (full dots) and Mg/Zr/Co (empty dots) multilayers.

The analysis of the Co Lα emission is more complicated. First, it must be kept in mind that the emission energy, 776.4 eV(Jonnard & Bonnelle, 2011), is only 2.4 eV away from the Co $2p_{3/2}$ ionization threshold, 778.8 eV(Jonnard & Bonnelle, 2011). That is to say, the Co optical constants used in the simulations have to be chosen in a range where they are strongly varying and also not known with high accuracy. This introduced an uncertainty on the fit of the angular curves. This is in contrast to the case of the Mg Kα emission, where the emission energy, 1253.6 eV(Jonnard & Bonnelle, 2011) is about 50 eV lower than the Mg 1s ionization threshold, 1303.4 eV(Jonnard & Bonnelle, 2011).

For the Co Lα emission of the Mg/Co/Zr system, the simulations reproduce the general shape of the GI and GE curves. These simulations are done with starting parameters deduced from the reflectivity measurements, the optical indices of the CXRO database (http://henke.lbl.gov/optical_constants/) and assuming a stack without inter-diffusion at the interfaces, that is to say by using a tri-layer stack. For the Mg/Zr/Co system, this description fails even to reproduce the contrast of the features, see Fig. 5c and 5d. Keeping a tri-layer structure but allowing the variation of the Co optical constants, the agreement between experimental and calculated emission curves is still not satisfying. In consequence, we modify Mg/Zr/Co as a Mg/Co$_x$Zr$_y$ bi-layer stack where $x$ and $y$ are the relative numbers of Co and Zr atoms, respectively. The values of $x$ and $y$, estimated from the number of Co and Zr atoms introduced within the Co and Zr layers in Mg/Zr/Co during its deposition, are found equal to 0.78 and 0.22, respectively, so that x/y is equal to 3.5. Such values are close to the ones of the Co$_{23}$Zr$_6$ compound present in the Zr–Co binary phase diagram(Predel, 1991). The simulations performed by modelling the stack as a Mg/Co$_x$Zr$_y$ bi-layer stack greatly improve the agreement with the experiment, see Fig. 5c and 5d. A bi-layer stack of the same composition was already used to fit the extreme UV reflectivity curves of a similar system, however with much thicker Mg layers (13 nm), designed to work around 25 nm(Le Guen et al., 2011a). Let us note that regarding the Mg Kα curve, there is no significant difference between the simulations with a tri- or bi-layer system, as the Mg atoms are not involved in interfacial diffusion.

Thus, the drastic differences between the curves of Mg/Co/Zr and Mg/Zr/Co, allow us to deduce that the Zr-on-Co interfaces in the Mg/Co/Zr multilayer are abrupt, whereas the Co-on-Zr interfaces in the Mg/Zr/Co multilayer are wide enough to consider the two Co and Zr layers as a single layer. Generally, alternate interfaces in multilayer structure are asymmetric due to the difference in the surface free energy of the constituents of the multilayer. This is analogous to what happens in the particularly well documented for the Mo/Si system(Yulin et al., 2002).

In order to understand the interfacial behaviour of the Mg/Zr/Co system, we calculated the mixing enthalpies of the Co-Mg, Co-Zr and Mg-Zr systems by using the Miedema's "macroscopic atom" model, whose details are given elsewhere(Miedema et al., 1980;

Das et al., 2013). This approach already proved useful to explain the interfacial phenomena taking place in annealed Co/Mo$_2$C nanometer multilayers(Yuan et al., 2015). The parameters used for the calculations, the electronegativities $\emptyset^*$, the electron densities at the first Wigner-Seitz boundary $n_{ws}$ and the atomic volumes V, are collected in Table 2.

**Table 2.** Elemental values of $\emptyset^*$, $n_{ws}$ and $V$ for calculating the mixing enthalpies of the Co-Mg, Co-Zr, Zr-Mg systems.

| Element | $\emptyset^*$(V) | $n_{ws}^{1/3}$(d.u.)$^{1/3}$ | V($cm^3/mol$) |
|---|---|---|---|
| Mg | 3.42 | 1.20 | 13.97 |
| Co | 5.10 | 1.77 | 6.70 |
| Zr | 3.62 | 1.38 | 14.10 |

We show in Figure 7 the calculated mixing enthalpies as a function of the mole fraction. The Co-Zr system shows a negative mixing enthalpy over the whole composition range, whereas the Co-Mg and Mg-Zr systems both show only positive values. This indicates that a compound is easy to form at the interfaces between the Co and Zr layers, which is not possible at the interfaces involving the Mg layers. Moreover, there is a large difference in the surface free energies of Co and Zr, 2.0 J.m$^{-2}$ and 1.6 J.m$^{-2}$ respectively(Vitos et al., 1998). Thus, during the deposition of the Co atoms on the Zr layers, since the surface free energy of Zr is lower, Co atoms can move on the surface guided by the chemical driving force. This leads to a strong mixing at the interfaces and to the formation of the Co$_x$Zr$_y$ compound in the Mg/Zr/Co multilayer. On the other hand, in the Mg/Co/Zr multilayer, during the deposition of the Zr atoms onto the Co layers, no such chemical driving force exists. Therefore the intermixing at Zr-on-Co interface takes place as the result of random thermal motions only and hence the concentration profile is expected to be an abrupt error function.This is consistent with previous results(Le Guen et al., 2011a; b; Zhu et al., 2011) on large period samples, demonstrating the good optical performances of Mg/Co/Zr owing to abrupt interfaces, whereas it was shown that Mg/Zr/Co has to be considered as a bi-layer system Mg/Co$_x$Zr$_y$.

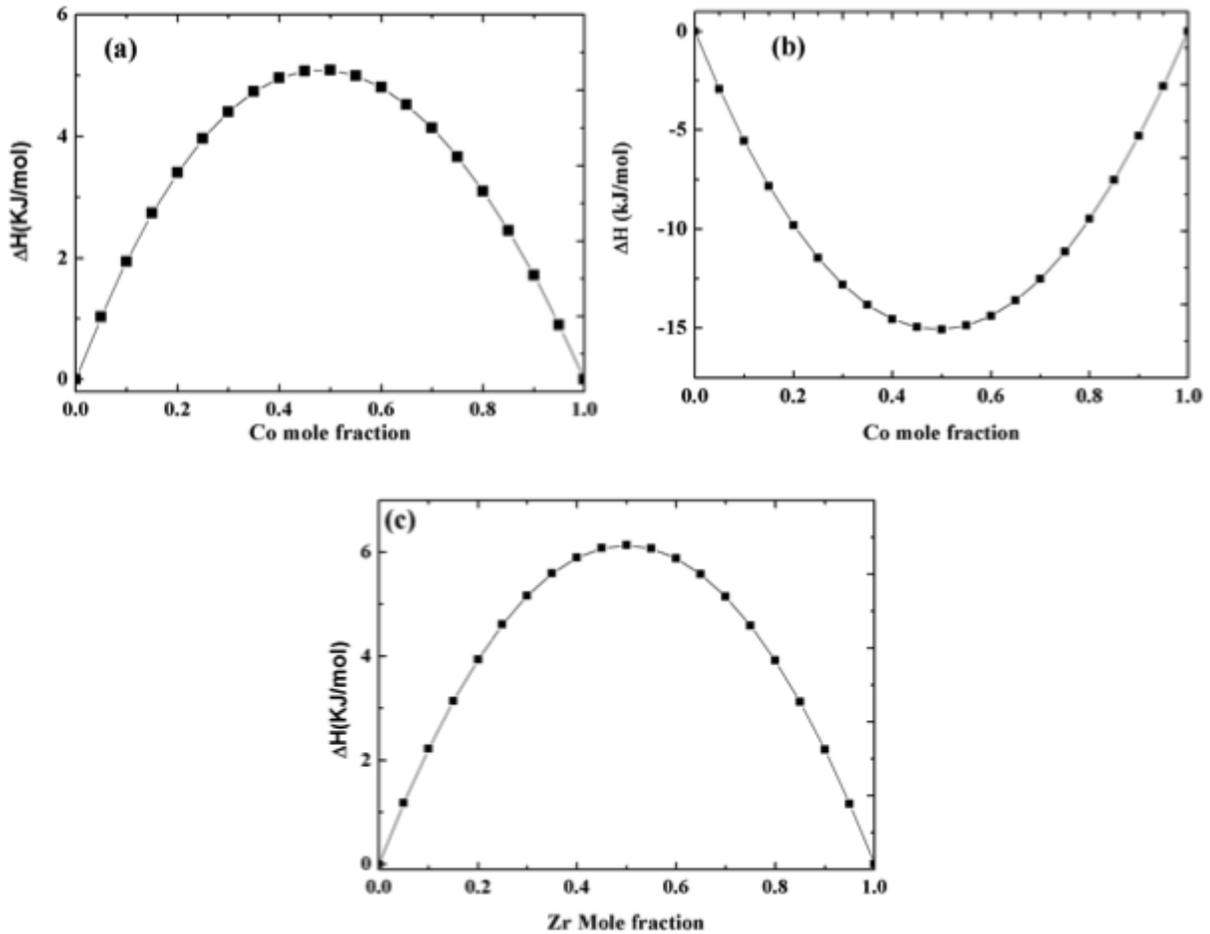

**Figure 7.** Calculated mixing enthalpies of the Co-Mg (a), Co-Zr (b) and Zr-Mg (c) systems.

4. **CONCLUSION**

Whereas the reflectivity measurement in the hard x-ray range was not enough sensitive to detect the asymmetric behaviour of interfaces in periodic multilayers, XRF generated by soft x-ray XSW, in particular after the improvement of the angular resolution, prove useful to study the structure of periodic multilayer stacks and to characterize their buried interfaces in a non-destructive way. The combination of both GI and GE modes enables to put higher constraints on the fitting parameters of the XSW curves. Working in the soft x-ray range gives sensitivity to the chemical state of the atoms owing to the large variation of the absorption coefficient in the vicinity of an absorption edge. Thus the contrast of the XSW curves depends on the chemical bound of the emitting element. This enables to couple the depth sensitivity of the XSW technique to the elemental and chemical sensitivity of XRF. Also, let us remark, that photon excitation is not mandatory in the GE mode. In this case, XSW intensity measurements can be also made upon electron or ion irradiation, thus do not require a synchrotron facility.

The combination of GI- and GE-XRF generated by XSW allows us to determine that Mg/Co/Zr and Mg/Zr/Co multilayers have to be considered as tri-layer and bi-layer stacks respectively, owing to the asymmetric behaviour of the Zr-on-Co and Co-on-Zr interfaces following the positive mixing enthalpy of the Co-Zr system and the different values of the surface free energies of Co and Zr. Thus the Mg/Zr/Co multilayer can be described as $Mg/Co_xZr_y$ with *x/y* around 3.5.

ACKNOWLEDGMENT

results has received funding from the European Community's Seventh Framework Programme (FP7/2007-2013) CALIPSO under grant agreement n°312284.